\documentclass[12pt]{article}
\usepackage{graphics,cite,amssymb,epsfig,float,psfrag}
\usepackage[usenames,dvips]{color}
\usepackage{rotating}

\begin{document}

\def\cO#1{{\cal{O}}\left(#1\right)}
\newcommand{\be}{\begin{equation}}
\newcommand{\ee}{\end{equation}}
\newcommand{\br}{\begin{eqnarray}}
\newcommand{\er}{\end{eqnarray}}
\newcommand{\ba}{\begin{array}}
\newcommand{\ea}{\end{array}}
\newcommand{\bi}{\begin{itemize}}
\newcommand{\ei}{\end{itemize}}
\newcommand{\bn}{\begin{enumerate}}
\newcommand{\en}{\end{enumerate}}
\newcommand{\bc}{\begin{center}}
\newcommand{\ec}{\end{center}}
\newcommand{\non}{\nonumbers}
\newcommand{\ul}{\underline}
\newcommand{\ol}{\overline}
\newcommand{\ra}{\rightarrow}
\newcommand{\sm}{${\cal {SM}}$}
\newcommand{\as}{\alpha_s}
\newcommand{\aem}{\alpha_{em}}
\newcommand{\ycut}{y_{\mathrm{cut}}}
\newcommand{\susy}{{{SUSY}}}
\newcommand{\Dir}{\kern -6.4pt\Big{/}}
\newcommand{\Dirin}{\kern -10.4pt\Big{/}\kern 4.4pt}
\newcommand{\DDir}{\kern -10.6pt\Big{/}}
\newcommand{\DGir}{\kern -6.0pt\Big{/}}
\newcommand{\gl}{\tilde g}
\newcommand{\sq}{\tilde q}
\newcommand{\ch}{\tilde \chi^\pm}
\newcommand{\NO}{\tilde\chi^0}
\newcommand{\stau}{\tilde\tau}
\newcommand{\st}{\tilde t}
\def\Ecm{\ifmmode{E_{\mathrm{cm}}}\else{$E_{\mathrm{cm}}$}\fi}
\def\gluino{\ifmmode{\mathaccent"7E g}\else{$\mathaccent"7E g$}\fi}
\def\photino{\ifmmode{\mathaccent"7E \gamma}\else{$\mathaccent"7E \gamma$}\fi}
\def\gl{\ifmmode{m_{\mathaccent"7E g}}
             \else{$m_{\mathaccent"7E g}$}\fi}
\def\taugluino{\ifmmode{\tau_{\mathaccent"7E g}}
             \else{$\tau_{\mathaccent"7E g}$}\fi}
\def\mphotino{\ifmmode{m_{\mathaccent"7E \gamma}}
             \else{$m_{\mathaccent"7E \gamma}$}\fi}
\def\ML{\ifmmode{{\mathaccent"7E M}_L}
             \else{${\mathaccent"7E M}_L$}\fi}
\def\MR{\ifmmode{{\mathaccent"7E M}_R}
             \else{${\mathaccent"7E M}_R$}\fi}
\def\lsim{\buildrel{\scriptscriptstyle <}\over{\scriptscriptstyle\sim}}
\def\gsim{\buildrel{\scriptscriptstyle >}\over{\scriptscriptstyle\sim}}
\def\jp #1 #2 #3 {{J.~Phys.} {#1} (#2) #3}
\def\pl #1 #2 #3 {{Phys.~Lett.} {#1} (#2) #3}
\def\np #1 #2 #3 {{Nucl.~Phys.} {#1} (#2) #3}
\def\zp #1 #2 #3 {{Z.~Phys.} {#1} (#2) #3}
\def\pr #1 #2 #3 {{Phys.~Rev.} {#1} (#2) #3}
\def\prep #1 #2 #3 {{Phys.~Rep.} {#1} (#2) #3}
\def\prl #1 #2 #3 {{Phys.~Rev.~Lett.} {#1} (#2) #3}
\def\mpl #1 #2 #3 {{Mod.~Phys.~Lett.} {#1} (#2) #3}
\def\rmp #1 #2 #3 {{Rev. Mod. Phys.} {#1} (#2) #3}
\def\sjnp #1 #2 #3 {{Sov. J. Nucl. Phys.} {#1} (#2) #3}
\def\cpc #1 #2 #3 {{Comp. Phys. Comm.} {#1} (#2) #3}
\def\xx #1 #2 #3 {{#1}, (#2) #3}
\def\NP(#1,#2,#3){Nucl.\ Phys.\ \issue(#1,#2,#3)}
\def\PL(#1,#2,#3){Phys.\ Lett.\ \issue(#1,#2,#3)}
\def\PRD(#1,#2,#3){Phys.\ Rev.\ D \issue(#1,#2,#3)}
\def\preprint{{preprint}}
\def\Ord{\lower .7ex\hbox{$\;\stackrel{\textstyle <}{\sim}\;$}}
\def\OOrd{\lower .7ex\hbox{$\;\stackrel{\textstyle >}{\sim}\;$}}
\def \MCH {$\tilde\chi_1^+$}
\def \CH{{\tilde\chi}^{\pm}}
\def \N0{\tilde\chi^0}
\def \LSP{\tilde\chi_1^0}
\def \SNU{\tilde{\nu}}
\def \BARSNU{\tilde{\bar{\nu}}}
\def \MLSP{m_{{\tilde\chi_1}^0}}
\def \MCH{m_{{\tilde\chi}^{\pm}}}
\def \MCHMIN {\MCH^{min}}
\def \ET{\not\!\!{E_T}}
\def \LL{\tilde{l}_L}
\def \LR{\tilde{l}_R}
\def \MLL{m_{\tilde{l}_L}}
\def \MLR{m_{\tilde{l}_R}}
\def \MSNU{m_{\tilde{\nu}}}
\def \PROCESS{e^+e^- \rightarrow \tilde{\chi}^+ \tilde{\chi}^- \gamma}
\def \PI{{\pi^{\pm}}}
\def \DM{{\Delta{m}}}
\newcommand{\bQ}{\overline{Q}}
\newcommand{\ad}{\dot{\alpha }}
\newcommand{\bd}{\dot{\beta }}
\newcommand{\dd}{\dot{\delta }}
\def \sq{\tilde q}
\def \gl{\tilde g}
\def \LSP{\tilde\chi_1^0}
\def \MUL{m_{\tilde{u}_L}}
\def \MUR{m_{\tilde{u}_R}}
\def \MDL{m_{\tilde{d}_L}}
\def \MDR{m_{\tilde{d}_R}}
\def \MSNU{m_{\tilde{\nu}}}
\def \MTAUL{m_{\tilde{\tau}_L}}
\def \MTAUR{m_{\tilde{\tau}_R}}
\def \mhf{m_{1/2}}
\def \MST{m_{\tilde t_1}}
\def \CHM{H^\pm}
\def \RPVC{\lambda'}
\def\tth{\tilde{t}\tilde{t}h}
\def\qqh{\tilde{q}_i \tilde{q}_i h}
\def\t1{\tilde t_1}
\def \ta1{\tilde\tau_1}
\def \MET{p{\!\!\!/}_T}  
\def \invfb{fb^{-1}}
\def\bul{\bullet}
\def \mht{H{\!\!\!/}_T}
\def\lapp{\mathrel{\rlap{\raise.5ex\hbox{$<$}}
                    {\lower.5ex\hbox{$\sim$}}}}
\def\gapp{\mathrel{\rlap{\raise.5ex\hbox{$>$}}
                    {\lower.5ex\hbox{$\sim$}}}}
\baselineskip=17pt
\begin{titlepage}
\begin{center}
{\sc\Large\bf Probing Supersymmetry using Event Shape variables 
at 8 TeV LHC}\\[5mm]
\vspace{0.5cm}
{\sc Rajdeep M. Chatterjee, Monoranjan Guchait, Dipan Sengupta}\\[2mm] 
{\noindent
Department of High Energy Physics,\\
Tata Institute of Fundamental Research,\\
1, Homi Bhabha Road, Mumbai 400 005\\
India.}
\\[4 mm]
\end{center} 

\vskip 50 pt 

\begin{abstract} 

\noindent 
\begin{quotation}
We have revisited the prospects of Supersymmetry(SUSY) searches at
the LHC with 7 TeV energy along with the prediction of the 
discovery potential at 8~TeV energy assuming an integrated 
luminosity 5 $fb ^{-1}$ and  20 $\invfb$
with
mSUGRA/CMSSM as a model framework.
We discuss further optimization of our selection strategy which is  
based on the hadronic event shape variables. Evaluating the standard 
model backgrounds and signal rates in detail we predict the
discovery reach in the $m_0 - m_{1/2}$ plane for  7~TeV with 
5$\invfb$ luminosity. We also present the discovery reach for 8~TeV 
energy with an integrated luminosity 5$\invfb$ and  20 $\invfb$.
A comparison is made between our results and the exclusion plots obtained by 
CMS and ATLAS. Finally, discovery reach in the gluino and squark mass plane 
at the 7~TeV and 8~TeV energy is also presented.
\end{quotation}
\end{abstract}
\vskip 60 pt
\begin{center} 
\end{center}
\begin{center}
\end{center}
\hrule
\vskip -10pt
\footnotesize \noindent
\centerline{
rajdeep@tifr.res.in \qquad
guchait@tifr.res.in \qquad
dipan@tifr.res.in} 
\normalsize
\end{titlepage}
\section{Introduction}
\label{sec_intro}
The Large Hadron Collider(LHC) at CERN has completed its run at  
7 TeV center of mass energy accumulating about 5$\invfb$ 
of data. While the early run focused primarily on reproducing the 
Standard model(SM) physics, it is heartening to see that as of now 
a huge amount of beyond standard model(BSM) physics has been probed, 
especially in the context of supersymmetry (SUSY).   
\vspace{0.1cm}

Supersymmetry over the last three decades has emerged as a robust 
and leading candidate for BSM. It is rich in phenomenological signatures 
and is a major search program at the LHC. The experiments CMS and ATLAS 
have both published results with $\sim$4.5$\invfb$ of data excluding a 
substantial region of parameter space from negative searches~\cite{cms,atlas} 
assuming one of the popular SUSY model, namely, constrained minimal 
suspersymmetric standard model(CMSSM) or minimal supergravity 
model(mSUGRA)~\cite{msugra}. For example, the current exclusion limit 
for gluino and squark mass is,
$\rm {m_{\gl}},\rm {m_{\sq}}\gsim$~1.2~TeV for 
$\rm {m_{\gl}} \sim \rm {m_{\sq}}$ case, 
while $\rm {m_{\gl}}\gsim $800 GeV
for $ \rm {m_{\sq}} >>\rm {m_{\gl}}$ case~\cite{cms,atlas}.   
\vspace{0.1cm}

However, the process of encompassing a wider region of
parameter space at a particular luminosity 
has been a constant quest for experimental and phenomenological 
study. Hence various methods have been devised to improve the
signal sensitivity by suppressing the backgrounds as much as possible, 
since the processes that lead to SUSY signals have a minuscule cross section
compared to the SM backgrounds.For instance, search analysis based on 
$\alpha_{T}$~\cite{alphat} variable is particularly
impressive in suppressing QCD background for dijet 
and as well as for multijet channels. In addition to this, the Razor variables 
~\cite{rogan} adopted by CMS collaboration use a different  
selection of strategy to detect SUSY signals. Following these examples,
we had tried to devise a new search strategy reported in 
Ref.~\cite{eventshape,nugmdm} which is proved to be robust for a vast 
region of parameter space in various SUSY models. We observed that 
this method is extremely effective for the regions of SUSY parameter 
space which yield events with higher multiplicity of hard jets.
In our paper~\cite{eventshape}, we demonstrated the 
robustness of our proposed selection strategy for 7 TeV energy by  
analyzing the signal and SM background events in details with 
1$\rm fb^{-1}$ integrated luminosity.
\vspace{0.1cm}

However, in this current study we revisited our search 
strategy with a goal for further optimization to make this 
method more powerful. After scrutinizing the previous selection strategy and 
correlation of cuts~\cite{eventshape} very closely, we found redundancy of
one of our cuts 
which is finally dropped from the present analysis without 
losing any signal sensitivity. In addition, we also update our study 
for 8 TeV LHC energy. In order to find the 
sensitivity of our search strategy we scan the entire region of mSUGRA/CMSSM 
parameter space and predict the discovery reach of SUSY signal.
Moreover, we also compare our results with that from CMS and ATLAS
for a given energy and luminosity.
\vspace{0.1cm}

We organize our paper as follows: In Sec.~\ref{sec:sigbg}, discussing our
selection variables we describe the search strategy. 
In Sec.~\ref{sec:res}, we present our results 
and summarize it in Sec.~\ref{sec:summary}. 
\section{Signal and Background}
\label{sec:sigbg}
In this current study we present our results assuming the mSUGRA/CMSSM as 
a model framework, which has been an extremely popular model of SUSY 
breaking for close to three decades~\cite{msugra}. Its attractiveness 
and popularity derives from the fact that it requires least number of 
assumptions about the parameters at the Grand Unified(GUT) scale. One 
assumes that at the GUT scale there is a universal scalar mass 
$\rm m_{0}$, a universal fermion mass $\rm m_{1/2}$, and a universal 
trilinear coupling $\rm A_{0}$. 
Along with these one has to provide the sign of the higgsino mass 
parameter $\rm \mu$ and tan$\rm \beta$, the ratio of the vacuum 
expectation values of the two Higgs doublet, both of  which are determined 
at the electro-weak(EW) scale. The entire spectrum at the EW scale is then 
generated as one runs down from the GUT scale to the EW scale via 
Renormalization Group Equations(RGE). The phenomenology at low energy is 
dependent on the couplings between various SUSY particles which drive the 
collider search strategies.
\vspace{0.1cm}

The mSUGRA/CMSSM is constrained by theoretical considerations like 
convergence of RGE's, perturbative unitarity constraining 
tan$\rm \beta$~\cite{djouadi}, as well as from non observation 
of SUSY particles in direct searches in collider experiments,
like at LEP \cite{pdg} and LHC~\cite{cms,atlas}. It is assumed to 
preserve the discrete symmetry R-parity and provides 
lightest neutralino as a dark matter candidate which is assumed to be the 
lightest SUSY particle(LSP). The observed dark matter relic density in 
the universe along with low energy constraints arising from flavor physics 
restrict the mSUGRA/CMSSM parameter space to a large extent which has 
been reported in a large number of papers~\cite{cmssmcon}. Interestingly, 
in a recent analysis based on very latest results from flavour data,
 in particular from LHCb measurements on $B_s \to \mu^+\mu^-$~\cite{lhcb}, 
it has been shown  that the mSUGRA/CMSSM is constrained severely, 
particularly for large $\tan\beta$ scenario~\cite{bsmumu}. Moreover, one 
of the most striking constraints may arise from the observation or 
non-observation of the light higgs boson which, as is indicative from 
the current data may well lie in the window 122-127 
GeV~\cite{cmshiggs,atlashiggs}, and will eventually rule out a large 
swath of the model parameter space. Again a number of phenomenological 
studies in the context of constraining the CMSSM parameter 
space \cite{susyhiggs} has been carried out in this light and will 
provide directions to future of SUSY. However, for a direct search 
of SUSY in colliders, it is better to remain unbiased from the indirect 
constraints and perform an inclusive search in all regions of parameter 
space except the regions which are forbidden  
from theoretical considerations and the direct search 
limits from collider experiments~\cite{pdg}.    
\vspace{0.1cm}

At the LHC the primary production modes of SUSY particles are
squark and gluino pairs mediated by 
strong interaction initiated by quarks and gluons
in the initial state. These gluinos and squarks being massive decay 
immediately into heavier chargino and neutralino states which further
decay to give rise jets and leptons along with lightest neutralinos 
which are LSPs and being undetectable experimentally lead to an 
imbalance in momentum at the   
transverse plane. Hence the generic SUSY signature is often designated 
by  {\it n-jets+ m-leptons+$\MET$} (n,m=0,1..). However, because of the 
depletion of branching ratio of SUSY particles in leptonic channels, 
the signal rates in the pure hadronic decay channels are much larger. 
Needless to say, that the jets + $\rm \MET$ channel is expected to yield 
the largest discovery reach than others. 
In this current effort we focus primarily on this jets+$\MET$ channel to 
investigate discovery potential of SUSY at the current runs of the LHC.
The major SM backgrounds corresponding to this final state 
consist of QCD, top pair production($t\bar t$+jets), W/Z+jets with
the hadronic decays of SM particles. As we know, the background
cross sections are larger by 6 or 7 orders of magnitude in 
comparison to signal cross sections, and hence  it is a non 
trivial task to isolate signal events from the rubble of these 
backgrounds. As mentioned above, in a previous work we formulated a new 
strategy~\cite{eventshape} with its own merit  to eliminate 
backgrounds to a large extent, particularly for the case where the 
signal has the characteristic feature of a larger multiplicity of hard jets. 
\vspace{0.1cm}

We use PYTHIA6~\cite{pythia} to simulate signal and
background  processes due to  $t\bar{t}$, QCD. The top
pair production and QCD is simulated by slicing the entire 
phase  space in $\hat p_{T}${\footnote {$\hat p_T$ is the transverse momentum
of final state partons in the center of mass frame.}} 
bins. We use CTEQ6L~\cite{cteq} for
parton distribution function to calculate cross sections setting 
the factorization scale at $Q^{2}=\hat{s}$, where $\sqrt{s}$ is the center 
of mass energy in the partonic frame. Jets are reconstructed using 
FastJet~\cite{fastjet} with an $ \rm {anti-K_{T}}$~\cite{ankt} 
algorithm using a size parameter R=0.5. Jets    
are selected with a cut of  $p_{T}^{j} \ge 50$GeV and 
$|\eta|\le3$. ALPGEN~\cite{alpgen} is used to generate 
events for  $t\bar{t}$+jets, W/Z+jets, applying
$p_{T}\ge 20$~GeV, $|\eta|\le 3$ for final state partons.
Subsequently, PYTHIA6~\cite{pythia} is used for parton showering
imposing MLM matching~\cite{mlm} with a jet $p_{T}>$25~GeV,
 $|\eta| \le 2.5 $ and $\Delta R=0.7 $ to avoid double counting.   
\vspace{0.1cm}

We pre-select events consisting jets and missing energy by imposing
the following selection: 
\br
p_T^j > 50~{\rm GeV}, |\eta|<3\ \  {\rm and}\ \  \MET>50~{\rm GeV}.
\label{eq:precuts}
\er
\vspace{0.1cm}
In the light of current experimental data where the mSUGRA/CMSSM is 
pushed to higher values of gluino and squark masses we choose two 
benchmark points in the  $m_{0}$ and $m_{1/2}$ plane 
consistent with current collider constraints \cite{cms,atlas} to
demonstrate our search strategy. In Table 1 we present the
masses of relevant SUSY particles corresponding to these 
parameters space P1 and P2. We use Suspect interfacing with 
SUSYHIT~\cite{shit} to generate masses and branching ratios of 
SUSY particles corresponding to these benchmark points. The SUSY particle 
production cross sections are calculated at next-to-leading(NLO) level using 
PROSPINO~\cite{prospino}.
\begin{table}
\begin{center}
\begin{tabular}{|c|c|c|c|c|c|c|c|c|c|c|c|c|}
\hline
Model&$\gl$ & $\sq$  & $\st_1$  & $\st_2$ & $\tilde{b}_1$
 &$\tilde{b}_2$&
$\tilde{e}_l$&$\tilde{\tau_{1}}$&$\tilde{\chi}_{1}^{0}$&
$\tilde{\chi}_{2}^{0}$
&$\tilde{\chi}_{1}^{+} $& $\tilde{\chi}_{2}^{+}$ \\
\hline
P1 & 825 & 1609 & 1009 & 1164 & 1155 & 1257 & 1507 & 1086
& 127 & 228 & 227& 354 \\
\hline 
P2 & 1217 & 1238 & 916  &  1150 & 1209  & 1198 & 707 & 641
 & 215 & 406 &  406 & 663 \\
\hline 
\end{tabular} 
\caption{
Mass spectrum for benchmark point  (P1) $m_0=$1500~GeV,
$m_{1/2}$=310~GeV tan$\beta$=10.$A_{0}$=0, sgn($\mu$)$\ge$0, 
(P2)$m_0=620$~GeV, $m_{1/2}=$520~GeV, tan$\beta$=10.
$A_{0}$=0, sgn($\mu$)$\ge$0
}
\end{center}
\label{table:spectrum}
\end{table}
In the following section we recapitulate our selection 
variables~\cite{eventshape} very briefly.
\vspace{0.1cm}

$\bullet$ \underline{Transverse Thrust(T) :} The event shape
variables describe the shape and geometry of the 
event~\cite{salam} and are used to understand various properties of it.
For instance, these variables are studied to test few 
monte carlo models at the LHC~\cite{evtlhc}. 
One of the event shape variable is the transverse thrust which 
is defined as~\cite{salam},
\br  
{\rm T} = {\rm max}_{n_T}\frac { \sum_i |\vec q_{T,i} . \vec n_T| }
{\sum_i |q_{T,i}| },
\label{eq:tht}
\er
where the sum runs over all objects in the event
with $\vec q_{T,i}$ being the transverse momentum component of each
object and $\vec n_T$ is the transverse unit vector which 
maximizes this ratio.
It is easy to convince oneself that
for di-jet events the ratio is precisely 1 and for multijets  
T is away from unity. Hence this provides an useful 
tool to suppress backgrounds with comparatively lower multiplicity of 
jets in the final state, predominantly for QCD and with some moderate 
effects on $t\bar t$ and W/Z+jets channels.  
This is clearly demonstrated in Fig.1 of Ref~\cite{eventshape}.  
A pleasing feature of transverse thrust is in the fact that it is infra-red 
safe~\cite{salam} and being a dimensionless quantity is expected to have 
less systematic errors. In practice, the transverse thrust is re-defined 
as $\rm\tau$=1-T.
\vspace{0.1cm}

$\bullet$ \underline{$R_{T}$ :} The variable $R_{T}$ is 
defined as~\cite{eventshape},
 \br
R_{T}(n_j^{min}) = \frac{\sum_1^{n_{j}^{min}} |\vec p_T^{j_i}|}{H_T}
\label{eq:RT}
\er
with $H_T = \sum_{1}^{n_j} |\vec p_{T}^{j_i}|$.
The numerator runs up to a required minimum number of selected  
jets($n_j^{min}$) of the event, whereas for the denominator it is the 
total number of jets existing in that event. It is evident that the   
pattern of distribution of this variable as was pointed out 
in Ref.\cite{eventshape}, depends
on the multiplicity and hardness of jets. SUSY as we noted earlier is
often characterized by hard multijet events in a wide region of parameter 
space and in addition, the sub-leading jets, 
are comparatively harder than that of the backgrounds. We make use of 
this distinguishing feature to separate  the signal
from the backgrounds using this variable. For instance, events where the total 
number of jets($n_j$) in the event is close to the required minimum 
number of jets($n_j^{min}$), 
i.e $n_j^{min}\sim n_j$, then the numerator and denominator are almost 
identical, hence the ratio is close to unity, which is predominant in 
backgrounds as can be seen in 
the Fig.2 of Ref~\cite{eventshape}. For signal processes, mostly 
$n_j^{min} << n_j$ with harder multijets, the ratio turns out to be away 
from 1 and the tail extends to well below 0.8. One can see that if we 
discriminate at $R_T(4)<0.85$, a significant fraction of backgrounds 
can be eliminated with a little effect on the signal cross section. 
\vspace{0.1cm}

$\bullet$ \underline{$M_{T}({j_{1},j_{2}})$} : We define
this variable as the transverse mass between the two
leading jets~\cite{eventshape},
\br
M_{T}({j_{1},j_{2}}) =
 \sqrt{2\times p_{T}^{j_{1}}p_{T}^{j_{2}}(1-\cos\phi(j_1,j_2))}
\label{eq:mT}
\er
\begin{figure}
\centering
\includegraphics[scale=0.8]{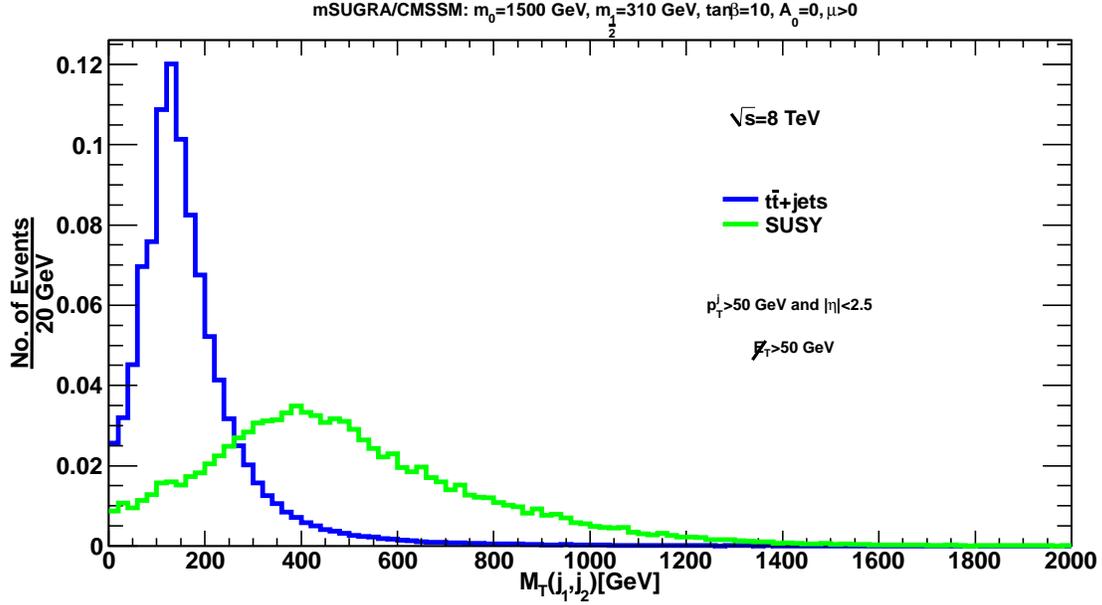}
\caption{Comparison of $M_{T}({j_{1},j_{2}})$ for $t \bar t$ and SUSY signal
events subject to pre-selection cuts(Eq.~\ref{eq:precuts}) and with
$\sqrt{S}=$8~TeV.  
Parameter space P1 is used for SUSY.}
\label{fig:MT}
\end{figure}
where $\phi$ is the azimuthal angle between the two leading jets.
This variable is particularly useful in suppressing the top background
where two leading jets are expected to be back-to-back mostly. On contrary,
in SUSY events the two leading jets are hard and very likely 
to be distributed isotropically as compared to the background. 
Fig.~\ref{fig:MT} clearly confirms this fact where we display the distribution
of $M_{T}({j_{1},j_{2}})$ for signal and $t \bar t$ background ignoring other
SM backgrounds which are manageable without this cut.    
For this illustration we use P1 parameter space to obtain signal distribution
and both are subject to
pre-selection cuts, Eq.~\ref{eq:precuts}. Clearly, a strong cut, for example,
$M_{T}({j_{1},j_{2}})>$450~GeV can bring down $t\bar t$ background  
to a negligible level. 
\vspace{0.1cm}

$\bullet$ \underline{$\MET$}: We calculate the total transverse missing 
momentum out of all visible particles
with $p_T>$~1 GeV and $|\eta|<$3. In SUSY process $\MET$ is expected
to be hard.
\vspace{0.1cm}

A careful comparison of all distributions of $\tau, R_T(n_j^{min}),
M_T(j_1,j_2)$ and $\MET$ for signal and backgrounds, we optimize the following 
set of cuts to reject backgrounds and isolate signal events,
\bc
\br
{\rm \tau} > 0.1,\ \  {\rm R_T(4)}<0.85,  \nonumber \\ 
{\rm M_T(j_1,j_2)}>450~{\rm GeV}, \MET > 250~{\rm GeV}
\label{eq:rejcuts}
\er 
\ec
In the next section we will discuss the impact of these cuts in both 
signal and background processes.
Meanwhile, we make some comments about the correlation of cuts
observed in our analysis,  
\vspace{0.1cm}

In SUSY events the $H_{T}$ distribution in signal 
is expected to be on the higher side as jets emerging from 
cascade decays of heavier particles are more 
energetic than their SM counterparts and hence, it 
is used as one of the background rejection tool.  
Following this observation, in our previous analysis we also adopted 
this $H_T$ variable to eliminate SM backgrounds~\cite{eventshape}. 
However, in this present study we investigate the correlation 
of $H_T$ variable with other cuts in more detail. 
\begin{figure}
\centering
\includegraphics[scale=0.8]{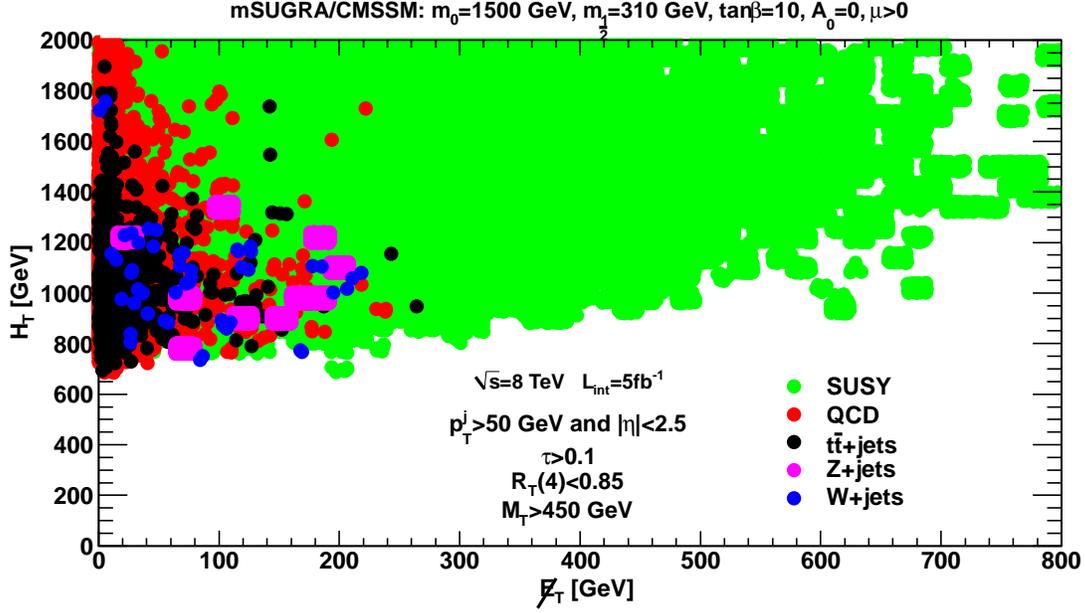}
\caption{Distribution of signal and background events with 
$\rm \tau>0.1$, $R_{T}(4)<0.85$, $M_{T}^{j_1,j_2}>450$~GeV
along with pre-selection cuts, Eq.\ref{eq:precuts}.
Parameter space P1 is used for SUSY.
}
\label{fig:metht}
\end{figure}
In Fig.~\ref{fig:metht}, we present the interplay of cuts 
by making a two dimensional plot in $\MET - H_T$ 
plane imposing selection on $\tau$ and $R_T(4)$, as Eq.~\ref{eq:rejcuts},
requiring at least 4 jets in the event for both signal and all SM backgrounds.
The $\MET - H_T$ distribution shown in Fig.~\ref{fig:metht} clearly 
reveals that signal events are located at the high $H_T$ 
($\gsim$750 GeV) region and in addition, requiring $\MET>$250~GeV 
it is possible to get rid of contamination due to the SM backgrounds.
This exercise justifies the claim of dropping $H_T$ cut from our 
selection strategy. These features  are  in stark contrast
to almost all multijet search strategies which requires a 
large $H_{T}$ cut in their analysis. One of the most important background 
to SUSY searches is the irreducible
$Z(\rightarrow~ \nu\bar{\nu})$+jets background with multijets
and a large amount of $\MET$~\cite{mangano}. 
Using $R_T(4)$ we have successfully managed to suppress this background 
to a negligible level. It must also be emphasized that this variable being a
dimensionless quantity is prone to less systematics,
and is a fairly simple variable to implement in experiments.
\section{Results and discussion}
\label{sec:res}
In order to understand the impact of our selection strategy as discussed 
in the previous section, 
we analyzed SUSY signal for two benchmark points shown in Table  
1. One of the points(P1) corresponds to a lower 
gluino mass and a relatively higher squark mass in contrast to
the other point(P2) where gluino and squark 
masses are almost equal. 
\begin{table}
\begin{center}
\begin{tabular}{|c|c|c|c|c|c|c|c||}
\hline
Process&  C.S(pb) & N & $\tau$ & $R_T{4}$ & $M_{T}({j_{1}j_{2})}$ &
 $ \MET $ & \# of Events  \\
 & & & $>0.1$   &$\le .85$  & $\ge 450 $ GeV &$ \ge 250 ~$ GeV  
& ${\cal L}=5\invfb$ \\
\hline
 P1 & &&&&&&\\
\hline
$\tilde{g}\tilde{g}$ &0.087 & 20K & 16809 & 9186 & 3840 & 1025 &
 22.3\\
\hline
$\tilde{q}\tilde{g}$ & 0.023 & 20K & 16474  & 9776 & 7363 & 3458 &
 19\\
\hline
P2 & &&&&&&\\
\hline
$\tilde{g}\tilde{g}$ & 0.002 & 20k & 17781 & 13650 & 6227 & 3810 &
 1.73\\
\hline
$\tilde{q}\tilde{g}$ & 0.015 & 20K & 14895 & 5286 & 3490 & 2883 & 
10.5\\
\hline
$\tilde{q}\tilde{q}$ & 0.02 & 20k & 10713 & 1068 & 451  & 299 & 
1.3\\ 
\hline
$t\bar{t}$ &&&& &   &  & \\
\hline
5-200 & 85 & 0.3M & 147181 & 5738 & 133 & 0 & 0 \\
\hline
200-500 & 10 & 0.1M & 29490 & 4518 & 328 & 2 & 1 \\
\hline
500-inf & 0.13 & 20k & 1986 & 248 & 147 & 9 & 0.3 \\
\hline
$t\bar{t}+ 1 j$ & 79.6 & 136083 & 68854 & 3354 & 20 & 0 & 0 \\
\hline
$t\bar{t}+ 2j$ & 39.6 & 192983 & 11110 & 1180 & 14 & 0 & 0 \\
\hline
$t\bar{t}+ 3j$ & 14.7 & 14993 &  9802 & 2239 & 110 & 0 & 0 \\ 
\hline
$t\bar{t}+ 4j$ & 4.5 & 12439 & 9192 & 3724 & 433 & 6 & 1.6 \\ 
\hline
QCD &&&&&&& \\
\hline
300-500 & 1267 & 2M & 263823 & 11765 & 4409 & 0  & 0 \\
\hline
500-800 & 67 & 0.3M  & 32646 & 1720 & 1439 & 0 & 0 \\
\hline
800-1500 & 3  & 0.1M  & 8110 & 412 & 394 & 0 & 0 \\
\hline
1500-inf& 0.01 & 10k & 496 & 10 &10 & 0 & 0 \\
\hline
W+2j & 1665 & 220879  & 122079 & 2 &   0 & 0  & 0 \\
\hline
W+3 j & 436.2 & 99616 & 43712 &  3 & 0  & 0 & 0 \\
\hline
W+ 4j & 105.3 & 68923 & 25324  & 342 & 36 &0    & 0 \\
\hline
Z+2j  & 1670  &120199  &67406  &0   &0   &  0 & 0  \\
\hline
Z+3j  & 450 & 241202  & 106864 & 6 & 0 & 0  & 0 \\
\hline
Z+4j  & 110 & 39203 & 17706 & 133 & 10 & 0 & 0 \\
\hline 
\end{tabular}
\caption{
Number of events after each set of cuts for signal and background for
$\sqrt{S}=$8~TeV.
In the the last column, number of events are normalized for 5 $\rm fb^{-1}$
luminosity.
} 
\end{center}
\label{table:evtsm}
\end{table}
\vspace{0.1cm}
In Table~2, 
we summarize the cumulative effect of cuts for a center of 
mass energy 8~TeV. The production cross sections(CS) are presented
in the second column of Table~2. It is to be noted that
the signal cross sections are at the next-leading-order(NLO)~\cite{prospino} 
level where as partially higher order corrections are taken into 
account by considering associated jets 
in the background processes. The 3rd column shows the number of events(N) 
generated and it is made sure that it corresponds
to at least 5$\invfb$ integrated luminosity. From
the 4th column onwards the number of events due to cumulative
effect of cuts are presented. Finally, in the last column, the number 
of events selected after all cuts including 
matching efficiencies(Eq.~\ref{eq:rejcuts}) are 
shown normalized to cross section for 5$\invfb$ luminosity.
In addition we also simulate signal and backgrounds with proper statistics
for 20$\invfb$ luminosity for which only final results are presented.
\vspace{0.1cm}

The benchmark points(P1 and P2) are so chosen as to reveal the difference
in the type of event distribution for the two points. The 
first point P1( Table 1) has a lower gluino mass and a comparatively 
higher squark mass which implies that the primary decay mode 
of gluino will be through 
$\rm \tilde{g}\to t b \chi^{\pm}_{1},t \bar{t}\chi_{1,2}^{0}$
via virtual top squarks. With top decaying in the hadronic mode 
for about 2/3, it yields a large number of jets. 
We find as expected that the suppression due to the 
thrust cut($\tau>0.1$) for the signal is  
about 20\% whereas for background it is the close to 
90\% for some cases, QCD in particular.
The $\rm R_{T}(4)$ selection variable is effective for 
multijet backgrounds and is reflected in the 5th column of 
Table~2. Eventually, the $\rm M_{T}({j_1,j_2})$ cut as discussed 
previously is useful to get rid of the remaining top background.
However, in case of parameter space P1 i.e for
high $m_0$ and low $m_{1/2}$, the mass differences among 
$\rm \chi^{\pm}_{1},\N0_2$ and 
$\rm \chi^{0}_{1}$ are comparatively small resulting in less available 
energy for final state particles  
leading to a softer spectrum including soft $\MET$. As a consequence, the 
effect of $\MET$($>$250~GeV) cut is severe for signal in this case, 
as reflected in the penultimate column in Table~2. 
Hence, total acceptance efficiency turns out to be small 
yielding low signal sensitivity in this region. 
\vspace{0.1cm}

The benchmark point P2 with $m_0$ and $m_{1/2}$ nearly equal, is different 
in the fact that the gluinos will preferentially decay to 
$\rm \tilde{t}~t$ with physical top squarks decaying further 
to $t~ \chi^{0}_{1,2}$. Hence the gluino decay will 
still yield a fair number of jets in the final states. The 1st two 
generation squarks will however decay predominantly to 
$\rm q\chi^{\pm}_{1}$ with charginos decaying to $W\chi^{0}_{1}$. 
This channel therefore yields a less jet activity in most cases 
which is suppressed by the $\rm R_T(4)<0.85$, as can be seen from Table 2.
\begin{figure}
\centering
\includegraphics[scale=0.9]{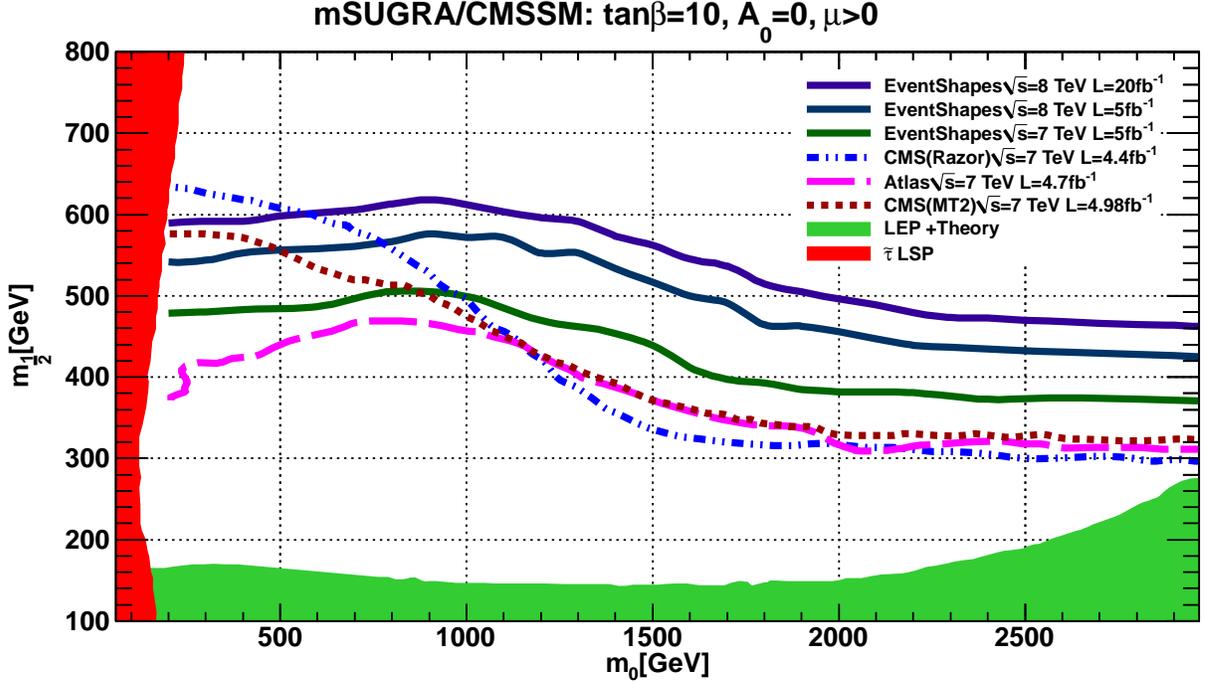}
\caption{Discovery reach requiring S$/\sqrt{B}\ge$5 for tan$\beta$= 10,
$\rm A_{0}=0,$ sign($\mu$)=+1. The two CMS(MT2 and Razor)~\cite{MT2,razor} and 
ATLAS~\cite{atlas} exclusion plots are at 95\% C.L. The green shaded 
region is disallowed
by theory and LEP constraints, red shaded region is forbidden
by $\tilde\tau_1$ LSP condition.
}
\label{fig:tanbeta10}
\end{figure}

\begin{figure}
\centering
\includegraphics[scale=0.9]{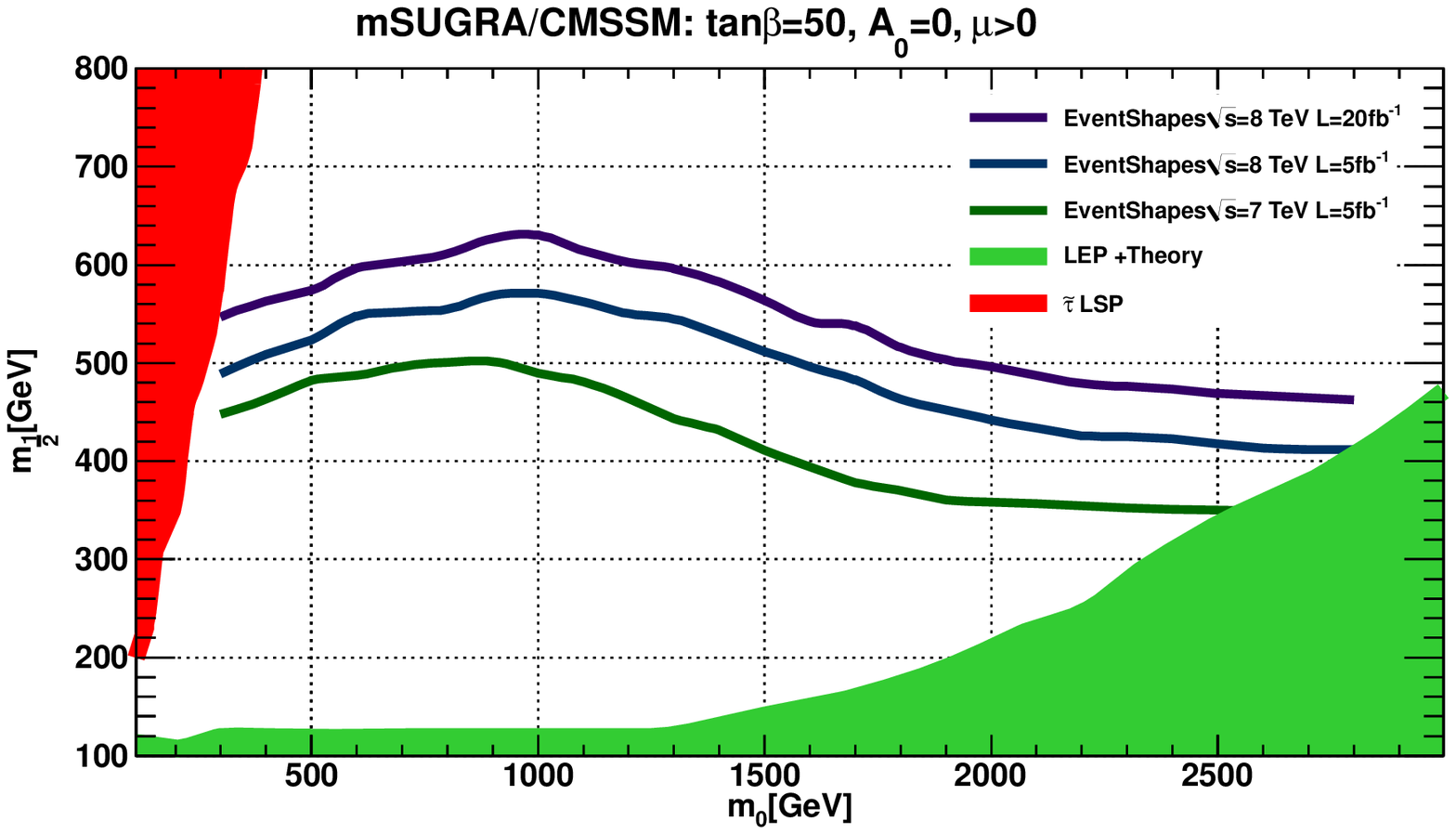}
\caption{
Same as Fig.~\ref{fig:tanbeta10}, but for tan$\beta$=50.
}
\label{fig:tanbeta50}
\end{figure}
\begin{table}
\begin{center}
\begin{tabular}{|c|c|c|c|c|c|c|}
\hline
$\sqrt{S}$ & $t \bar t$ & W+jets & Z+jets & QCD & Total Bg 
& SUSY \\
TeV & & & & & &P1~~~~~P2 \\
\hline
7 & 2.83 & $<$1 & 1.32 & $<1$ &4.15 & 20.7~~~~~5.24 \\
8 & 2.9 & $<$1 & $<1$ & $<1$ & 2.9 & 41.3~~~~~13.53 \\  
\hline
8 & 7.2 & $<$1 & $<1$ & $<1$ & 7.2 & 165~~~~~~54 \\ 
\hline
\end{tabular}
\caption{
First two rows(last row) present the number of signal and background events for 
5$\invfb$(20 $\invfb$) luminosity subject to all selection cuts, 
(Eq.~\ref{eq:precuts},\ref{eq:rejcuts}) corresponding to center of mass 
energies 
as shown. 
 } 
\end{center}
\label{table:evt5fb}
\end{table}
\vspace{0.1cm}
In Table~\ref{table:evt5fb}, we show the total number of background and
signal events after all selection cuts for two parameters points P1 and 
P2 normalizing to cross section at 5$\invfb$ luminosity. 
We observe about 4(3) background events for integrated 
luminosity 5$\invfb$ at 7(8)~TeV energy against a handful of signal 
events yielding $S$/$\sqrt{B}$ more than 
5 for two selected representative signal parameters points P1 and P2. 
The suppression of background events indicate the robustness of our 
selection strategy.
\vspace{0.1cm}

Armed with this selection strategy, we attempt to find the potential 
discovery region in the $m_0 - m_{1/2}$ plane.  
We scan the $m_0 - m_{1/2}$ parameter space setting $A_0$=0, sign($\mu$)=+1,
$\tan\beta$=10, 50 and estimate the signal rates
applying cuts, Eq.~\ref{eq:precuts} and \ref{eq:rejcuts}.  
We require $\rm{S/\sqrt{B}} \ge$5 to claim discovery of SUSY signal 
for each set of parameters points for a given energy and luminosity. In   
Fig.~\ref{fig:tanbeta10} and Fig.~\ref{fig:tanbeta50} 
we present the discovery reach in the $m_0 - m_{1/2}$ plane 
for $\tan\beta=10$ and $\tan\beta=50$ respectively.
In both the figures the shaded area along the x-axis are mainly 
disallowed by no-EWSB breaking condition as well as the limit on 
chargino mass($>$102~GeV) from LEP experiments~\cite{pdg}. On the other 
hand the shaded region along the y-axis are ruled out 
because $\tilde\tau_1$ appears to be LSP which is assumed to be forbidden 
because of offering LSP as a dark matter candidate which has to be neutral.
We present our results for 7 TeV with  
5 $\invfb$ luminosity   and 8 TeV  
energy with 5$\invfb$ and 20 $\invfb$ luminosity.
The total background and signal cross sections are presented in Table 3.
It is expected that the 
discovery reach for 8 TeV is higher and this enhancement occurs mainly due to
the enhancement of sparticle production 
cross sections, approximately by a factor of 2. Notice that
in the same plane we also delineate regions excluded at 95\% C.L.
by CMS and ATLAS at 7 TeV energy  
with 4.4$\invfb$ and 4.7$\invfb$ luminosity respectively. 
Note that for $\tan\beta=$50 case, Fig.~\ref{fig:tanbeta50}, 
exclusion plots are not available from both the experiments at 
this integrated luminosity. 
Notice that the two CMS exclusion plots are due to the two methods
MT2~\cite{MT2} and Razor~\cite{razor} with almost same luminosity. 
The ATLAS exclusion plot is obtained by demanding the number of jets $\ge$6 to 
$\ge$9 along with $\MET$ in the final states~\cite{atlas}, 
which is the similar type of final states where our search strategy 
is most sensitive.It is to be emphasized
that in both CMS and ATLAS analysis, no isolated leptons, electrons or muons 
are required. It helps to suppress backgrounds, mainly due to $t \bar t$ 
and W+jets. However, in our analysis we do not require to veto any such 
events to suppress these backgrounds. It seems from these figures 
that our selection strategy works better for high $m_0$ 
values where as for low $m_0$ case it is comparable with other results. 
A naive comparison of our results with a recent paper of Ref.~\cite{baer} 
which predicts 
the glunio mass up to $\sim$800~GeV whereas our analysis 
claims it $\sim$1 TeV 
for 7 TeV 5 $\rm fb^{-1}$ luminosity in the 
high $\rm m_{0}$($>$1500 GeV) region. It is to be noted that the signal rates
in the paper\cite{baer} correspond to inclusive channel, but in our case
it is  due to the jets plus $\MET$ channel. 
It is true that at the high $m_0$, as discussed before, the $\MET$ in 
the events is softer and hence signal selection based on tight cut 
on $\MET$ suffers and sensitivity degrades very fast. However, in 
our case, instead of high $\MET$ cut, we exploit the 
multiplicity of jets in the events, which is on higher side in this 
high $m_0$ region due to the presence of heavy flavors(t,b quarks) 
in $\gl,\sq$, cascade decay chains as discussed previously.
As a consequence, selections based on our strategy achieves better
significance than the others which are based on very hard cut on $\MET$ and 
$H_T$. On the other hand, towards the higher side of $m_{1/2}$ 
and comparatively low $m_0$ values, the masses of gluinos and 
squarks are close to each other,  
multiplicity of jets is relatively lower and hence our strategy suffers
to some extent.We observed that the difference in $\rm tan \beta$ does not 
make a significant impact in the discovery reach, which is
expected as rates in the hadronic channel is controlled by the cascade
decays of the strong production process whereas $\rm tan \beta$
affects the electroweak processes. However, the
differences are subtle 
and appear only in parts of parameter space which has 
$\tilde{\tau_1}$ as the next to LSP due to large tan $\rm \beta$ 
and hence yield $\tau$ leptons in the final state. This yields a lesser
number of jets in some parts of parameter space which
in our search strategy translates to a lower reach in large 
$\tan \beta$ region.
\begin{figure*}[t]
\centering
\includegraphics[scale=0.9]
{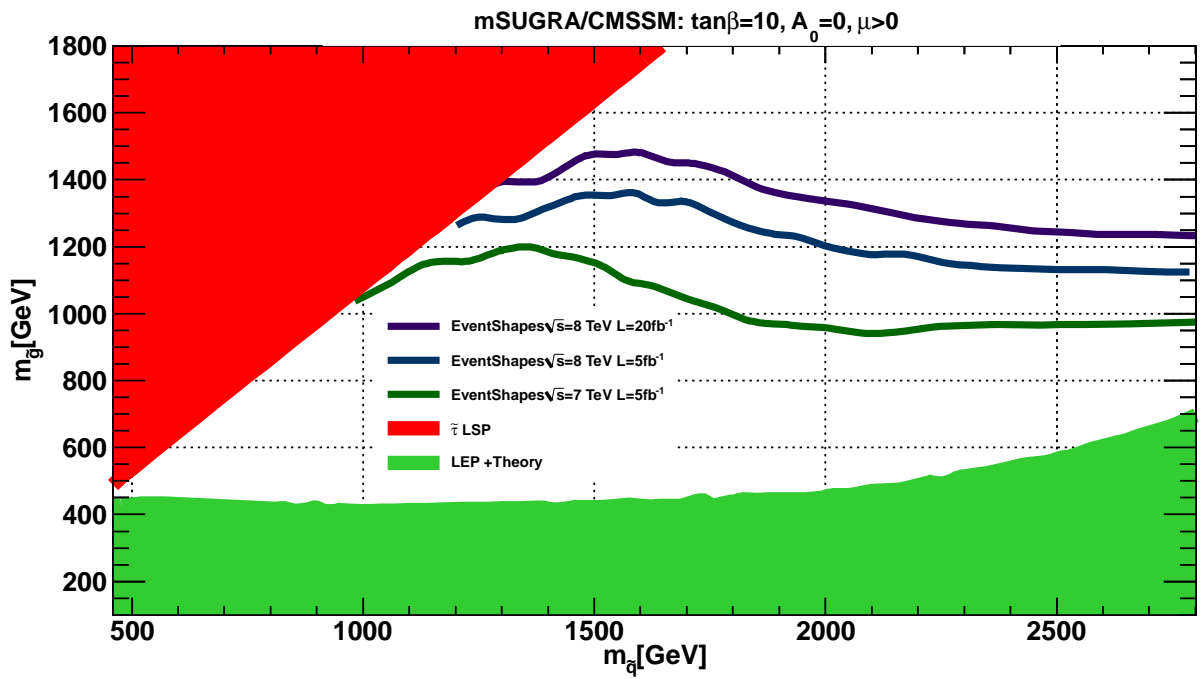}
\caption{Discovery reach for tan $\rm \beta$ = 10,
$\rm A_{0}=0,$ sign($\mu$=+1).
}
\label{fig:mglmsqtb10}
\end{figure*}

\begin{figure*}[t]
\centering
\includegraphics[scale=0.9]
{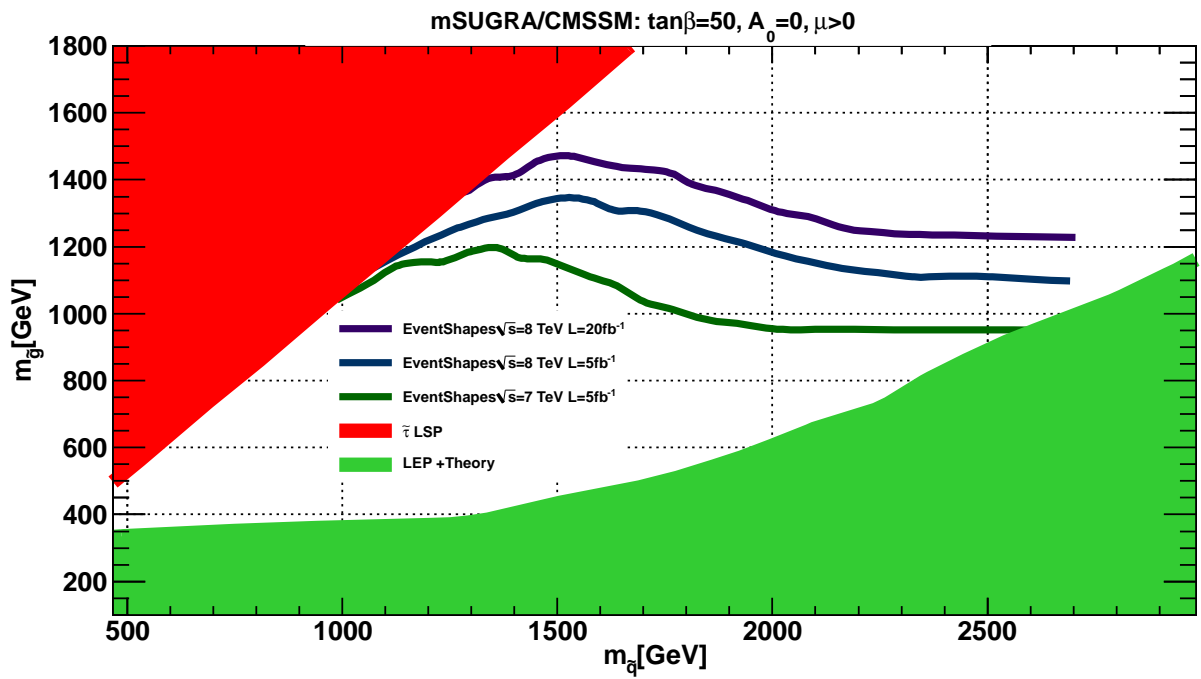}
\caption{Same as Fig.~\ref{fig:mglmsqtb10} but for tan$\beta$=50.
}
\label{fig:mglmsqtb50}
\end{figure*}
In order to understand the implication of this discovery region in 
$m_0 - m_{1/2}$ planes, we translate Figs.~\ref{fig:tanbeta10} 
and \ref{fig:tanbeta50} to Figs.~\ref{fig:mglmsqtb10} and \ref{fig:mglmsqtb50}
respectively, which are in the 
physical $m_{\tilde g} -  m_{\sq}$ mass planes.
Clearly, both the figures display the discovery reach of masses of $\gl$
for the corresponding $\sq$ masses and vice versa for a given set of
SUSY parameter space. We find that for nearly degeneracy case,
$m_{\tilde g} \sim  m_{\sq}$, it is possible to find SUSY 
signal for $m_{\tilde g}$ up to 1.2 TeV(1.35 TeV)
for 7~TeV(8 TeV) energy with 5$\invfb$ luminosity where as for 
larger masses of $\sq$, this reach goes down to 1 TeV(1.1 TeV)
for the same energy range. 
This conclusion remains true for high $\tan\beta$, 
(Fig.~\ref{fig:mglmsqtb50}) case as well.    
For higher luminosity options, say 20$\invfb$ 
for 8 TeV energy this reach extends to $\sim$1.5 TeV. 
Note that our predictions are based purely from generator level analysis
without taking care of any detector effects.

\section{Summary}
\label{sec:summary}
We have re-examined the prospects of SUSY searches in the jets plus missing 
energy channel at the LHC 
with 7 TeV energy and predicted the discovery reach for 8 TeV energy
with 5$\invfb$ and 20 $\invfb$ luminosity. 
We are familiar with the fact that the SUSY events are characterized 
by high multiplicity of comparatively harder jets for a wide 
region of parameter space. Exploiting this fact we devise our selection 
strategy based on the event shape variables, namely the transverse 
thrust(Eq.~\ref{eq:tht}) and the other observables, Eq.~\ref{eq:RT} 
and \ref{eq:mT}, which are robust in dealing with SM backgrounds. 
As mentioned in Sec.~\ref{sec:sigbg}, the $H_T$ cut is very effective 
and used by many SUSY analysis including by us in our previous 
study~\cite{eventshape}. However, investigating
the correlation of cuts used in the present analysis we observed that
due to selections given by Eq.~\ref{eq:rejcuts}, signal events are distributed 
to very higher $H_T$ region implying the inefficiency of $H_T$ cut, and 
hence it is dropped from our selection strategy without paying any price
for signal sensitivity.  
\vspace{0.1cm}

Moreover, optimizing our selection strategy,
we analyze the SM backgrounds in detail and showed 
that with our search strategy it is possible to reduce these to 
a negligible level retaining the signal to a large extent. Recall that, 
our search 
strategy works favorably in the regions of parameter space 
which predict events with many harder jets. We predict the 
potential discovery regions in the $m_0 - m_{1/2}$ plane 
in the framework of mSUGRA/CMSSM for 7 and 8 ~TeV centre 
of mass energy and 5 $\rm fb^{-1}$ and 20 $\rm \invfb$
luminosity. A naive comparison 
with the ATLAS, CMS
exclusion plots at 7 TeV suggests that our analysis gives 
better sensitivity in the regions of parameter space in 
high $\rm m_{0}$, and low  $m_{1/2}$ viz, in regions where jet activity is 
significantly higher. We find based on our analysis, $m_{\gl}$  
up to 1.35 TeV can be discovered for
gluino and squark mass degeneracy case and otherwise up to 1.1 TeV for  
large squark masses for 8 TeV energy. For higher luminosity options, 
for instance
${\cal L}=20\invfb$ case, this discovery reach is extended  
to about 1.45 TeV.
It must be noted that we have not vetoed any events
consisting of leptons or do not apply any tagging of any special objects, like
b-tag and so on. Since, two of our selection variables, namely 
$\tau$ and $R_T$, 
are a dimensionless quantities, so are expected to carry less systematic 
uncertainties.
\vspace{0.1cm}

As masses of SUSY particles in mSUGRA/CMSSM model gets pushed 
to a corner it is imperative to design 
search strategies that will access the edges of the SUSY 
parameter space. This will involve optimizing the signal to 
background ratio in large parts of the parameter space where 
the signal cross section is miniscule. In this study we have
provided such a search strategy
with its own merits of suppressing SM backgrounds to a negligible
level. It must be emphasized
that our strategy is not limited to mSUGRA/CMSSM but expected to work also in 
other models which yield a large number of jets, for instance 
non-universal gaugino mass model~\cite{nugmdm},
no scale F-SU(5)~\cite{noscale} SUSY model~\cite{progress}.

\end{document}